\documentclass[12pt,aps,prb,floatfix,showpacs,superscriptaddress]{revtex4}
\usepackage{amsmath}
\usepackage{citesort}

\begin{document}
\draft
\title{Microwave conductivity of YBa$_2$Cu$_3$O$_{6.99}$ including
inelastic scattering}
\author{E. Schachinger}
\affiliation{Institut f\"ur Theoretische Physik, Technische Universit\"at
Graz\\A-8010 Graz, Austria}
\author{J.P. Carbotte}
\affiliation{Department of Physics and Astronomy, McMaster University,\\
Hamilton, Ont. L8S 4M1, Canada}
\date{\today}
\begin{abstract}
The fluctuation spectrum responsible for the inelastic scattering
in YBa$_2$Cu$_3$O$_{6.99}$ which was recently determined from consideration
of the in-plane optical conductivity in the infrared, is used to
calculate the temperature dependence of the microwave conductivity
at several measured frequencies. Reasonable overall agreement can
only be achieved if, in addition, some impurity scattering is
included within a model potential intermediate between weak (Born)
and strong (unitary) limit.
\end{abstract}
\pacs{74.20.Mn 74.25.Gz 74.72.-h}
\maketitle
\newpage
\section{Introduction}

In a recent paper Carbotte {\it et al.}\cite{carb} showed that the
$41\,$meV spin resonance observed in spin polarized neutron
scattering experiments\cite{Bourges} on YBa$_2$Cu$_3$O$_{6.95}$
(YBCO) has a counterpart in the infrared optical conductivity
in which an optical resonance is seen at the same energy
demonstrating its coupling to the charge carriers.\cite{schach1,%
schach2} The temperature evolution of the spin resonance can be
determined from consideration of the frequency dependence of the
optical scattering rate measured at several different
temperatures\cite{schach3} in the superconducting state. The
required analysis which allows one to extract the resonance from
the optics is carried out within a generalized Eliashberg
formalism for a $d$-wave superconductor.\cite{schach4} The end
result is a charge carrier-exchange boson excitation spectral
density which characterizes the inelastic scattering. The spectral
density $I^2\chi(\omega)$ shows a distinct evolution with
decreasing temperature. At $T_c$ the spectrum obtained is linear in
$\omega$ at small $\omega$, followed by broad peak at some
characteristic energy $\omega_{SF}$ and then there is a very slow decay
at higher $\omega$ extending up to a cutoff $\omega_c$ of order
$400\,$meV. This spectrum can be fit with the spin fluctuation
form employed in Pines' group\cite{millis,mont}
$I^2{\omega/\omega_{SF}\over1+(\omega/\omega_{SF})^2}\theta(\omega_c-
\omega)$ (MMP model), where $I^2$ is a coupling constant to the charge
carriers. As the temperature is lowered below $T_c$, the
fluctuation spectrum obtained from the infrared data undergoes
important changes. It develops a gap at the lowest energies
as well as a peak at $41\,$meV. The strength of this peak tracks
well the observed growth of the area under the spin susceptibility,
$\Im{\rm m}\{\chi({\bf q},\omega)\}$ at ${\bf q} = (\pi,\pi)$,
measured by Dai {\it et al.}\cite{dai} in neutron scattering.
The higher energy part, however, remains largely unaffected. The
observed changes in spectral density are as expected and are interpreted
as coming from feedback effects on the excitation spectrum due to
the changes brought about in the electronic system by the onset
of superconductivity. This arises in any purely electronic
mechanism of superconductivity in which the pairing proceeds
through bosons exchanged between the charge carriers and where the
bosons themselves are intrinsic to the electronic system.%
\cite{schach4,nuss} The changes in the low energy part of
$I^2\chi(\omega)$ lead directly to the phenomenon referred to
as the collapse of the quasiparticle scattering rate and is
responsible for the
freezing out of the inelastic scattering at low temperatures.
In turn, this leads to a prominent peak around $30\,$K in the
in-plane microwave conductivity\cite{bonna} as a function of
temperature $T$ and a corresponding peak in the electronic thermal
conductivity.\cite{matsuk}

Schachinger {\it et al.}\cite{schach3} have recently found that
this same spectral density $I^2\chi(\omega)$ also gives, within
an Eliashberg formalism,
good agreement with observed properties of the superconducting
state such as the ratio of the gap to the critical temperature, the
fractional optical spectral weight involved in the condensation
(superfluid stiffness), the temperature dependence of the
penetration depth, the magnitude of the condensation energy,
as well as other quantities. It is also obvious from the
way the temperature dependent spectral density has been
derived in Ref.~5 that the coupling to the $41\,$meV spin
resonance in YBCO cannot play the role of the `glue' leading
to superconductivity\cite{Kee} because it is absent at $T_c$.

In this paper we consider the temperature dependence of the
microwave conductivity for five different frequencies
between 1 and $75\,$GHz observed recently\cite{hoss} in
ultra pure samples of YBa$_2$Cu$_3$O$_{6.99}$ grown in
BaZnO$_3$ crucibles. Our calculations are based on the
previously determined spectral density $I^2\chi(\omega)$
for twinned YBCO single crystals which is not modified in any
way. It is found, however, that to understand the low
temperature data ($10-20\,$K) it is necessary to additionally introduce
some elastic impurity scattering. A model impurity potential
intermediate between weak (Born) and strong (unitary) scattering
is developed which provides a reasonable, if not perfect overall
fit to the data.

In Sec.~II we give the necessary formalism. This is followed by
the presentation of our results in Sec.~III. Comparison with
the data is also presented in this section. Section IV contains
discussion and a conclusion.

\section{Formalism}

We consider a $d$-wave superconductor. To include the inelastic
scattering which is known to be strong in the cuprates we need
to go beyond a simple BCS formalism. The minimum set of equations
that allow us to do this are generalized Eliashberg equations.
These equations involve two channels. The ordinary renormalization
channel which remains in the normal state and leads to
renormalization of the Matsubara frequencies by the interactions.
The second is the pairing channel which we assume to have
$d$-wave character. For simplicity we take the pairing potential
to be separable in incoming $({\bf k})$ and outgoing $({\bf k}')$
momentum in the two dimensional CuO$_2$ Brillouin zone. The
charge carrier-fluctuation spectrum spectral density $I^2\chi(\omega)$
which characterizes the inelastic scattering in the renormalization
channel is assumed to also cause the pairing and while, in principle,
the spectral density could have a different shape in this case,
for simplicity, we ignore such complications. We do allow, however,
for a possible difference in magnitude through a constant $g$. For the
renormalized frequencies $\tilde{\omega}$ we use $I^2\chi(\omega)$
and $\tilde{\omega}(\omega+i\delta)$ is isotropic while for the
pairing energy $\tilde{\Delta}$ we use $gI^2\chi(\omega)\cos(2\phi)%
\cos(2\phi')$ which immediately leads to a $d$-wave form
$\tilde{\Delta}(\omega+i\delta)\sim\cos(2\phi)$, where $\phi$
is the angle defining the direction of momentum ${\bf k}$ on
a cylindrical Fermi surface.

The generalized Eliashberg equations which play the central role
in this study are
\begin{subequations}
\label{eq:1}
\begin{eqnarray}
  \tilde{\Delta}(\nu+i\delta;\phi) &=& \pi Tg
  \sum\limits_{m=0}^\infty\cos(2\phi)\left[\lambda(\nu-i\omega_m)+
  \lambda(\nu+i\omega_m)\right]\left\langle
  {\tilde{\Delta}(i\omega_m;\phi')\cos(2\phi')\over
  \sqrt{\tilde{\omega}^2(i\omega_m)+\tilde{\Delta}^2(i\omega_m;
  \phi')}}\right\rangle'\nonumber\\
 &&+i\pi g\int\limits^\infty_{-\infty}\!dz\,\cos(2\phi)
  I^2\chi(z)\left[n(z)+f(z-\nu)\right]\nonumber\\
 &&\times\left\langle
  {\tilde{\Delta}(\nu-z+i\delta;\phi')\cos(2\phi')\over
  \sqrt{\tilde{\omega}^2(\nu-z+i\delta)-\tilde{\Delta}^2(\nu-z+i\delta;
  \phi')}}\right\rangle',\label{eq:1a}
\end{eqnarray}
and in the renormalization channel
\begin{eqnarray}
  \tilde{\omega}(\nu+i\delta) &=& \nu+i\pi T%
  \sum\limits_{m=0}^\infty\left[\lambda(\nu-i\omega_m)-
  \lambda(\nu+i\omega_m)\right]\left\langle
  {\tilde{\omega}(i\omega_m)\over
  \sqrt{\tilde{\omega}^2(i\omega_m)+\tilde{\Delta}^2(i\omega_m;
  \phi')}}\right\rangle'\nonumber\\
  &&+i\pi\int\limits^\infty_{-\infty}\!dz\,
   I^2\chi(z)\left[n(z)+f(z-\nu)\right]\nonumber\\
  &&\times\left\langle
  {\tilde{\omega}(\nu-z+i\delta)\over
  \sqrt{\tilde{\omega}^2(\nu-z+i\delta)-\tilde{\Delta}^2(\nu-z+i\delta;
  \phi')}}\right\rangle' + i\pi\Gamma^+{\Omega(\nu)\over
   c^2+D^2(\nu)+\Omega^2(\nu)}
  \label{eq:1b}.
\end{eqnarray}
\end{subequations}
In the above $\tilde{\Delta}(i\omega_m;\phi)$ is the pairing energy
evaluated at the fermionic Matsubara frequencies
$\omega_m = \pi T(2m-1), m = 0,\pm1,\pm2,\ldots$. $f(z)$ and $n(z)$
are the Fermi and Bose distributions respectively. The renormalized
Matsubara frequencies are $\tilde{\omega}(i\omega_m)$. The
analytic continuation to real frequency $\nu$ of the above is
$\tilde{\Delta}(\nu+i\delta;\phi)$ and $\tilde{\omega}%
(\omega+i\delta)$, where $\delta$ is a positive infinitesimal.
The bracket $\langle\cdots\rangle$ is the angular average over $\phi$, and
\begin{equation}
  \label{eq:2}
  \lambda(\nu) = \int\limits^\infty_{-\infty}\!d\Omega\,
   {I^2\chi(\omega)\over\nu-\Omega+i0^+},
\end{equation}
\begin{equation}
  \label{eq:3}
  D(\nu) =  \left\langle{\tilde{\Delta}(\nu+i\delta;\phi)\over
  \sqrt{\tilde{\omega}^2(\nu+i\delta)-\tilde{\Delta}^2(\nu+i\delta;
  \phi)}}\right\rangle,
\end{equation}
\begin{equation}
  \label{eq:4}
  \Omega(\nu) = \left\langle{\tilde{\omega}(\nu+i\delta)\over
  \sqrt{\tilde{\omega}^2(\nu+i\delta)-\tilde{\Delta}^2(\nu+i\delta;
  \phi)}}\right\rangle.
\end{equation}
Equations (\ref{eq:1}) are a set of nonlinear
coupled equations for the renormalized pairing potential $\tilde{\Delta}%
(\nu+i\delta;\phi)$ and the renormalized frequencies
$\tilde{\omega}(\nu+i\delta)$ with the gap
\begin{equation}
  \label{eq:5}
  \Delta(\nu+i\delta;\phi) = {\tilde{\Delta}(\nu+i\delta;\phi)%
  \over Z(\nu)},  
\end{equation}
where the renormalization function $Z(\nu)$ was introduced in the
usual way as $\tilde{\omega}(\nu+i\delta) = \nu Z(\nu)$.
In Eq.~(\ref{eq:1b})
$\Gamma^+$ sets the size of the impurity scattering and $c$ is
a parameter related to the impurity potential. The Born or
weak scattering limit corresponds to a large value of $c$
while the unitary or strong scattering limit corresponds to
$c=0$. The impurity term is obtained from a T-matrix approach
to the impurity problem.\cite{proh} It does not include
all possible complications that have come to be known as
possibly of some importance in the cuprates.
Recent STM data\cite{pan,huds} has revealed significant inhomogeneities
and impurity studies based on BdG equations have shown that the
superconducting order parameter is strongly modified in the
vicinity of an impurity.\cite{frank,ghosal,franz,zhu1,atkinson1,%
atkinson2,balatsky,salkola,zhu2} Such complications go well
beyond the present approach. Here we will not attempt a
complete microscopic description of the impurity scattering
but instead treat $\Gamma^+$ and $c$ as parameters which we
will determine through a best fit to data. To get a better
fit may well require the introduction of more sophisticated
effective potentials.

In the pure limit, i.e.: no impurity contribution $(\Gamma^+ = 0)$
all parameters in Eqs.~(\ref{eq:1}) that serve to characterize
the particular material of interest, are fixed from our previous
work. The charge carrier-fluctuation spectrum spectral density
$I^2\chi(\omega)$ which enters through Eq.~(\ref{eq:2}) was
previously obtained by Schachinger {\it et al.}\cite{schach3}
from infrared optical data using an inversion technique\cite{mars1}
which allows one to construct 
$I^2\chi(\omega)$ from the optical scattering rate.
In the present calculations we simply use these results without
any modifications. The parameter $g$ is also fixed, and was
determined to get the measured value of the critical temperature
in YBCO. While we will first present results in the clean limit
we will later see that, to understand the low temperature data
$(T\to 0)$ we will need to consider some impurity scattering.
The value for $\Gamma^+$ and $c$ will be chosen to best fit the
data as we have already emphasized.

The optical conductivity follows from a knowledge of $\tilde{\omega}$
and $\tilde{\Delta}$. The formula to be evaluated is
\widetext
\begin{subequations}
\label{eq:sig}
\begin{eqnarray}
\lefteqn{\sigma_{ab}(\Omega) = {i\over\Omega}\frac{e^2N(0)v^2_F}
  {2}}\nonumber\\
 && \times\left\langle\int\limits_0^\infty\!{\rm d}\nu\,
  {\rm tanh}\left({\nu\over 2T}\right)\frac{1}
  {E(\nu;\phi)+E(\nu+\Omega;\phi)}\left[
  1-N(\nu;\phi)N(\nu+\Omega;\phi)-
  P(\nu;\phi)P(\nu+\Omega;\phi)\right]\right.
  \nonumber\\
 &&\left.+\int\limits_0^\infty\!{\rm d}\nu\,
  {\rm tanh}\left({\nu+\Omega\over 2T}\right)\frac{1}
  {E^\star(\nu;\phi)+E^\star(\nu+\Omega;\phi)}
  \left[1-N^\star(\nu;\phi) N^\star(\nu+\Omega;\phi)
  \right.\right.\nonumber\\
 &&\left.\left.
  -P^\star(\nu;\phi) P^\star(\nu+\Omega;\phi)
  \right]%\right.\nonumber\\
  +\int\limits_0^\infty\!{\rm d}\nu\,\left[{\rm tanh}
  \left({\nu+\Omega\over 2T}\right)-{\rm tanh}\left({
  \nu\over 2T}\right)\right]\right.\nonumber\\
 &&\left.\times\frac{1}
  {E(\nu+\Omega;\phi)-E^\star(\nu;\phi)}
  \left[1+N^\star(\nu;\phi) N(\nu+\Omega;\phi)
  +P^\star(\nu;\phi) P(\nu+\Omega;\phi)
  \right]\right.\nonumber\\
 &&\left.+\int\limits_{-\Omega}^0\!{\rm d}\nu\,
  {\rm tanh}\left({\nu+\Omega\over 2T}\right)\left\{
  \frac{1}
  {E^\star(\nu;\phi)+E^\star(\nu+\Omega;\phi)}
  \left[1-N^\star(\nu;\phi) N^\star(\nu+\Omega;\phi)
  \right.\right.\right.\nonumber\\
 &&\left.\left.\left.
   -P^\star(\nu;\phi) P^\star(\nu+\Omega;\phi)
  \right]\right.\right.\nonumber\\
 &&\left.\left.
   +\frac{1}
  {E(\nu+\Omega;\phi)-E^\star(\nu;\phi)}
  \left[1+N^\star(\nu;\phi) N(\nu+\Omega;\phi)
  +P^\star(\nu;\phi) P(\nu+\Omega;\phi)
  \right]\right\}\right\rangle  \label{eq:siga}
\end{eqnarray}
with
\begin{equation}
  E(\omega;\phi) = \sqrt{\tilde{\omega}^2_{\bf k}
   (\omega)-\tilde{\Delta}^2_{\bf k}
   (\omega)} \label{eq:sigb}
\end{equation}
and
\begin{equation}
  N(\omega;\phi) = {\tilde{\omega}_{\bf k}(\omega)\over
   E(\omega;\phi)},\qquad
  P(\omega;\phi) = {\tilde{\Delta}_{\bf k}(\omega)\over
   E(\omega;\phi)}. \label{eq:sigc}
\end{equation}
\end{subequations}
In the above, $\langle\cdots\rangle$ means, as before, an average
over the angle $\phi$ and the star refers to the complex conjugate.
$N(0)$ is the electronic density of states at the Fermi surface
and $v_F$ the Fermi velocity. The prefactor in (\ref{eq:siga}) can
be worked out to be proportional to the plasma frequency squared,
$\Omega^2_p/4\pi\equiv ne^2/m^\star$. Here $n$ is the electron density,
$e$ the electron charge, and $m^\star$ its effective mass.

\section{Results and Discussion}

In Fig.~\ref{f1} which has five frames we show our results
for the temperature dependence of the real part of the
microwave conductivity $\sigma_1(\omega)$ in the
pure limit (open triangles) i.e.: $\Gamma^+ = 0$ in Eq.~(\ref{eq:1})
at the five measured frequencies of Hosseini {\it et al.}\cite{hoss}
They correspond from top frame to bottom frame to $\Omega = 1.14$,
2.25, 13.4, 22.7, and $75.3\,$GHz respectively. For the lowest
frequencies considered, the agreement with the data (solid squares),
which we read of a graph in Ref.~11,
is not good but the agreement does improve as the microwave frequency
is increases. In particular, in the two top frames the
height of the calculated peak is too high and it falls at a somewhat lower
temperature than indicated in the measured curve. These deficiencies can
be largely removed when a small amount of impurity scattering is
additionally included. The effect of impurity scattering will
show up most prominently at the lowest temperature where the
inelastic scattering is becoming very small.

Before proceeding with a fit to the data which includes both impurities
and the inelastic scattering it is useful to first consider
the BCS limit of our generalized Eliashberg equations (\ref{eq:1})
and to understand the effect of impurities in this instance.
At low temperatures, the inelastic scattering rate which depends on real
processes is small and the impurities will dominate; thus the
BCS theory will become more applicable although it does ignore all
renormalizations from the inelastic interaction (virtual
processes).

To obtain the BCS equations from (\ref{eq:1}) we ignore the
effect of $I^2\chi(\omega)$ in the renormalization channel, so
\begin{equation}
  \label{eq:6}
  \tilde{\omega}(\nu+i\delta) = \nu+i\pi\Gamma^+{\Omega(\nu)%
   \over c^2+\Omega(\nu)^2}
\end{equation}
and in the gap channel we assume that the Boson frequency in
$I^2\chi(\omega)$ is very high compared with all other energies
of importance. This means that Bose and Fermi factors in the
second term on the right hand side of (\ref{eq:1a}) are negligible
and we can also replace the $\lambda(\nu\pm i\omega_m)$ by a constant
$(\lambda)$ with a
cutoff $(\omega_c)$ on the Matsubara sum applied to get convergence.
This gives a gap $\tilde{\Delta}(\phi)$ independent of frequency 
$\nu$ which satisfies the equation
\begin{equation}
  \label{eq:7}
  \tilde{\Delta}(\phi) = 2\pi T g\cos(2\phi)%
  \sum\limits_{m=0}^{\omega_c}\lambda\left\langle%
  {\tilde{\Delta}(\phi')\cos(2\phi')\over%
  \sqrt{\tilde{\omega}^2(i\omega_m)+\tilde{\Delta}^2(\phi')}}%
  \right\rangle'.
\end{equation}
The impurities enter directly in Eq.~(\ref{eq:6}) and affect the
gap given in Eq.~(\ref{eq:7}) through the renormalized Matsubara
frequency $\tilde{\omega}(i\omega_m)$ which appears in the square root
in the denominator. In the limit of $\nu = 0$ we can write
\begin{equation}
  \label{eq:8}
  \tilde{\omega}(\nu+i\delta) \approx i\gamma
  \equiv i\pi\Gamma^+{\Omega(i\gamma)%
   \over c^2+\Omega(i\gamma)^2}
\end{equation}
which can be solved self consistently for the impurity scattering
rate at zero frequency.
We can evaluate $\Omega(\tilde{\omega}) = 2K(\Delta_0/\tilde{\omega})%
/\pi$ where $K(x)$ is the elliptic integral of the first kind.\cite{hirsch1}
The quantity $\Omega(i\gamma)$ appearing in Eq.~(\ref{eq:8}) is
$\Omega(i\gamma) = [2\gamma/(\pi\Delta_0)]{\rm ln}(4\Delta_0/\gamma)$
and for a general value of $c$, the equation for $\gamma$ is
\begin{equation}
  \label{eq:9}
  \gamma = \pi\Gamma^+{{2\gamma\over\pi\Delta_0}{\rm ln}\left(
     {4\Delta_0\over\gamma}\right)\over c^2+\left(
     {2\gamma\over\pi\Delta_0}\right)^2{\rm ln}^2\left(
     {4\Delta_0\over\gamma}\right)}.
\end{equation}
This equation shows that the self consistent impurity scattering
rate $\gamma$ at zero frequency in the superconducting state is strongly
dependent on the parameter $c$. For the strong coupling unitary
limit $c=0$ an approximate solution has been given by
Hirschfeld and Goldenfeld\cite{hirsch1} for $\pi\Gamma^+\ll\Delta_0$
as
\begin{equation}
  \label{eq:10}
  \gamma \simeq 0.63\sqrt{\pi\Gamma^+\Delta_0}.
\end{equation}
Note that $\gamma(c=0)$ is much
larger than $\pi\Gamma^+$ in this limit. In the opposite limit
(Born limit or weak scattering potential) $c\to\infty$ and
$\pi\Gamma^+/c^2$ is to be replaced by $\pi\Gamma_N$ and
\begin{equation}
  \label{eq:11}
  \gamma(c\to\infty) = 4\Delta_0 e^{-\Delta_0/(2\Gamma_N)},
\end{equation}
which shows that $\gamma(c\to\infty)$ is now much smaller than
the normal state value of $\Gamma$, $\Gamma_N$.

This analysis demonstrates that the zero frequency 
self consistent scattering
rate in the superconducting state is much larger than its
normal state value in the unitary limit but is much smaller in
the Born limit. In particular, this implies that in the Born
limit the impurity limited quasiparticle mean free path
for a given impurity content will be much larger
in the superconducting state than in the corresponding normal
state if the inelastic scattering is ignored.

In Fig.~\ref{f2} we illustrate the general case. What is plotted
is $\gamma(c)$ as a function of $c$ for a specific value of
$\Gamma^+ = 0.15\,$meV and a zero temperature gap of
$\Delta_0 = 24\sqrt{2}\,$meV. The underlying normal state
scattering rate with which $\gamma(c)$ is to be compared is
$\pi\Gamma^+/(1+c^2)$ for any value of $c$. In this example
for $c=0$, $\gamma(c)$ is larger than $\pi\Gamma^+$ (by a
factor of 5), for $c=0.2$ they are comparable and for $c=0.3$
it is already much less. By changing $c$ we can change the
value of quasiparticle scattering rate at $\nu=0$ in the
superconducting state by orders of magnitude and this will be
of importance for our analysis of the experimental data.

It is instructive to look as well at the frequency dependence
of the underlying quasiparticle scattering rate, or more
precisely, the imaginary part of the renormalized frequency,
namely
\begin{equation}
  \label{eq:11a}
  \tau^{-1}(\nu) = \Im{\rm m}\tilde{\omega}(\nu) =
  \tilde{\omega}_2(\nu) = 
  \pi\Gamma^+{\Omega[\tilde{\omega}(\nu)]\over c^2+
  \Omega^2[\tilde{\omega}(\nu)]}.
\end{equation}
In Fig.~\ref{f3} we show results for several values of $c$.
The behavior of $\tau^{-1}(\nu)$ vs. $\nu$ at small $\nu$
changes radically with choice of $c$. In the unitary limit
there is a small region where $\tau^{-1}(\nu)$ is fairly
flat but for finite $c$, $\tau^{-1}(\nu)$ begins to look like
a $d$-wave quasiparticle density of states and the scattering 
is radically
affected by the onset of superconductivity which modifies the
density of final states available for scattering. These effects can
be understood simply in the clean limit $\Gamma^+\to 0$ and for
temperatures $T > \gamma$. This limit is considered in the
work of Hirschfeld {\it et al.}\cite{hirsch2} who treat
the two cases $c=0$ and $c\to\infty$ explicitely. Here we
consider finite $c$ and $\nu$ small $(\ll\Delta_0)$\cite{wu}
\begin{equation}
  \label{eq:12}
  \tau^{-1}(\nu) = \pi\Gamma^+{\nu\over\Delta_0}
  {c^2+A_+(\nu)\over c^4+2c^2A_-(\nu)+(4\nu/\Delta_0)^2
  A_+(\nu)}
\end{equation}
with
\begin{equation}
  \label{eq:13}
  A_\pm(\nu) = \left({4\nu\over\Delta_0}\right)^2\left[
  \left({\pi\over 2}\right)^2\pm{\rm ln}^2\left(
  {2\Delta_0\over\omega}\right)\right].
\end{equation}
It is clear that for $c\to\infty$ $\tau^{-1}(\nu)$ becomes
proportional to $\nu$ while for $c=0$ it goes like $\nu^{-1}$
at small $\nu$. For a general $c$, the quasiparticle
scattering rate $\tau^{-1}(\nu)$ is importantly
dependent on $\nu$ and, therefore, is quite different for the
constant of the familiar normal state Drude model. This means that,
while we have two parameters $\Gamma^+$ and $c$ to adjust, the
underlying complicated
variation of $\tau^{-1}(\nu)$ with $\nu$ gets reflected directly
in the frequency variation of the conductivity and leads to a
non-Drude form in sharp contrast to the underlying normal state.

We present results in Fig.~\ref{f4}. What is plotted is the
real part of the conductivity $\sigma_1(\omega,T)$ in units
of $N(0)v_F^2$ as a function of frequency $\omega$ in the
range up to $1.0\,$meV ($\sim 242\,$Ghz). The temperature is
set at $T=10\,$K and the impurity scattering at
$\Gamma^+ = 0.15\,$meV in Eq.~(\ref{eq:6}). Various values
of the impurity parameter $c$ are shown. The solid gray curve
in Fig.~\ref{f4} is the normal state shown for comparison.
It displays the classical Drude form with a Drude width of
$\pi\Gamma^+$. The other curves are in the superconducting
state at $T=10\,$K with the zero temperature $d$-wave gap
taken to be $\Delta_0 = 24\sqrt{2}\,$meV. The black solid
curve is the
Born limit, dotted the unitary limit, dashed $c=0.4$ and
dash-dotted $c=1.0$. None of the curves in the superconducting
state follow the Drude form of the normal state
$\sigma_1(\omega) = 2\tau_{imp}/[1+(\omega\tau_{imp})^2]$
with $\tau_{imp} = 2.12\,$meV$^{-1}$. The curves near the
Born limit show a concave up rather than the concave down
behavior of the normal state Drude. The curve for the unitary
limit is much flatter than the solid gray curve reflecting a
value of $\gamma(c)$ which is large compared with its normal
state counterpart. The inset in the top right hand shows this
on a different scale and allows the reader to better see the
radical difference in behavior between Born and unitary limit.
Although, neither of these limits show a Drude variation
with $\omega$, we can still think of the half width
of each curve as giving a measure of the underlying 
quasiparticle scattering rate. One then concludes that in
the Born limit it
is much smaller than in the normal state while in the
unitary limit it is much larger.

The data in the work of Hosseini {\it et al.}\cite{hoss} which
we use here for comparison with theory were fit to a Drude
form and the authors concluded that the effective quasiparticle
scattering rate was fairly temperature independent and constant.
It is clear from our Eq.~(\ref{eq:12}) and Fig.~\ref{f3} that
this behavior will never be reproduced in a BCS theory with
only elastic impurity scattering whatever the value of $c$.
The underlying quasiparticle scattering rates are always highly frequency
dependent and this modulates the Drude line
width as $\omega$ changes. This also implies a temperature dependence
since a change in $T$ involves a different sampling of the
frequency dependence of $\tau^{-1}(\omega)$. Any two fluid approach
would need to account for these features of the scattering rates
of the normal quasiparticles as well as the energy dependence
in the density of states.\cite{hirsch2} Including
some impurity scattering in addition to the inelastic scattering in
Eq.~(\ref{eq:1}) greatly improves the agreement with experiment.
The half width of the conductivity as a function of $\omega$
in our clean limit
Eliashberg calculations is significantly smaller than what is
observed experimentally. (See Fig.~\ref{f5}, open symbols.)
This does allow us to add some impurities which of course always result
in an increase of the half width.

To make an appropriate choice of $\Gamma^+$ and of $c$ we are guided by the
Drude analysis of the data provided by Hosseini {\it et al.}\cite{hoss}
They find a very narrow width to their Drude of the order of
$1/3\,$K. It is clear that the amount of impurities involved
is very small and that $\pi\Gamma^+/(1+c^2) = \gamma_N(c)$
which gives the scattering in the normal state is correspondingly small.
It is also clear that the unitary limit is unlikely since
this increases $\gamma(c)$ considerably as compared with the normal state
equivalent $\gamma_N(c)$ which would then have to be much
smaller than $1/3\,$K which is not likely. On the other hand,
the Born limit gives a concave upward curve which drops too
rapidly as $\omega$ increases out of zero. It appears that
some intermediate $c$-value is favored but on examination
of Fig.~\ref{f3} and the form (\ref{eq:12}) it is clear that
our best fit still differs from a Drude form for the conductivity.

After some trial and error we came up with $\Gamma^+ = 0.003\,$%
meV and $c=0.2$. Our new results which include inelastic as well
as impurity scattering are shown as the solid triangles in
Fig.~\ref{f1}. The agreement with the data is clearly greatly
improved over the pure limit, i.e.: $\Gamma^+ = 0$, in
Eq.~(\ref{eq:1}), particularly at low temperatures and for the
smaller microwave frequencies used in the experiment. The over all fit
to the entire data set is quite good but certainly not excellent.
The theoretical calculations do reproduce well the general trends
such as the decrease in peak height with increasing microwave
frequency and its shift to higher energies. The inelastic scattering
largely controls this trend and the present analysis provides support
for the validity of the charge carrier-fluctuation spectrum
spectral weight $I^2\chi(\omega)$ used here. We stress that the
form of $I^2\chi(\omega)$ was not adjusted in any way to fit the
microwave data but comes from consideration of the infrared
optical scattering rate only. 

One can further examine the quality of the fit by plotting
the same data as a function of frequency for fixed temperature $T$.
This is done in Fig.~\ref{f5}. In the figure the black solid symbols
are experiment, gray solid theory with impurities ($\Gamma^+ =%
0.003\,$meV, $c=0.2$) and the open symbols the pure case, i.e.:
including only the inelastic scattering captured in our model spectral
density $I^2\chi(\omega)$. It is quite clear that some impurity
scattering is needed to get even reasonably close to the data
although a tight fit is never possible. The data do not vary as
rapidly at the lowest frequency values as in theory. We point out, however,
that the fit is very much better than we could achieve using a
pure BCS formalism. Inclusion of inelastic scattering is essential
in any serious attempt to understand this data even at reasonably
low temperatures. Our use of the generalized Eliashberg equations,
given in the previous section, can be viewed as a phenomenology
with kernel $I^2\chi(\omega)$ determined from experimental data.
We have previously found that this phenomenology is able to
explain many of the anomalous superconducting properties observed
in the oxides. The present study extends the range of agreement to the
microwave data although some discrepancies do remain. These are not
large, however. In Fig.~\ref{f5} the data for $20\,$K fit perfectly,
for 10K we have only disagreement for 1 GHz and at 15 K we
have a slight disagreement for 1 and 2 GHz. But for these two
temperatures Hosseini {\it et al.}\cite{hoss} also have problems with
their Drude fits. The over all fit seems to be as
good as the one of Hosseini {\it et al.}\cite{hoss} and also as in the
analysis of Berlinsky {\it et al.}\cite{berlin} who conclude
that the data does not support a quasiparticle picture. Here
we find, instead, no serious disagreement of the data with
an Eliashberg formulation of the $d$-wave state which includes
some impurity scattering described with an intermediate value
of $c$, the parameter that spans the interaction strength
from unitary $(c=0)$ to Born $(c=\infty)$.

\section{Conclusion}

We have made use of a set of charge carrier-boson exchange
spectral densities $I^2\chi(\omega)$ obtained previously from
an analysis of infrared optical conductivity data to calculate the
microwave conductivity at several frequencies as a function
of temperature in a generalized Eliashberg formalism suitable
to describe a $d$-wave superconductor. Agreement with recent data
on pure samples
of YBa$_2$Cu$_3$O$_{6.99}$ is satisfactory provided a small
amount of elastic impurity scattering is also included. The
impurity potential used is neither in the Born (weak) nor
unitary (strong) limit. A potential of intermediate strength
is indicated. The low temperature behavior found in the
theory cannot accurately be described by a Drude form and does not
support the use of a two fluid model with the normal component
described by a scattering rate constant in frequency
although temperature dependent.
Instead, the conductivity as a function of energy 
at fixed $T$ is concave upward reflecting the intrinsic frequency
dependence of the combined scattering rates.
This holds even when inelastic scattering is
included. The calculations show clearly that, on the hold,
the main features of the microwave data can be understood
within the same generalized Eliashberg formalism that has
recently been so successful in describing many of the
anomalous superconducting state properties\cite{schach3}
seen in the oxides. This follows also from the observation
by Schachinger and Carbotte\cite{schach5} that adding
elastic impurity scattering only affects the low energy
region of the optical properties while the inelastic
scattering effects are seen in the energy region
$\omega > \Delta_0$. Thus, adding elastic impurity
scattering allows a fit of theoretical Eliashberg
results to match low energy optical properties of a
particular sample without violating all earlier findings which
particularly concentrated on the energy region
$\omega > \Delta_0$ or on bulk effects.

\section*{Acknowledgment}

Research supported by the Natural Sciences and Engineering
Research Counsel of Canada (NSERC) and by the Canadian
Institute for Advanced Research (CIAR).

\newpage
\begin{figure}
\caption{Microwave conductivity $\sigma_1(\omega,t)$ in
$10^7\Omega^{-1}$m$^{-1}$ vs. the reduced temperature
$t=T/T_c$ for the five frequencies measured in experimental
work of Hosseini {\it et al.}\cite{hoss} namely $\Omega =%
1.14\,$GHz, 2.25,13.4, 22.7, and $75.3\,$GHz (bottom frame).
Solid squares are experiment, open triangles clean limit and
solid triangles inelastic scattering plus impurities characterized
by a potential with $\Gamma^+ = 0.003\,$meV and $c=0.2$.}
\label{f1}
\end{figure}
\begin{figure}
\caption{The self consistent impurity scattering rate $\gamma(c)$
in the superconducting $d$-wave state given by Eq.~(\ref{eq:9}) for
various values of the impurity potential $c$. It is largest for
$c=0$ which corresponds to the unitary limit and rapidly becomes
small as $c$ increases beyond 0.3. The corresponding normal scattering
rate that we should compare $\gamma(c)$ with is
$\pi\Gamma^+/(1+c^2)$. Here, $\Gamma^+ = 0.15\,$meV and
$\Delta_0 = 24\sqrt{2}\,$meV.}
\label{f2}
\end{figure}
\begin{figure}
\caption{The imaginary part of the renormalized frequency
$\tilde{\omega}_2(\nu)$ as a function of $\nu$ for several
values of the impurity parameter $c$, namely $c=0$ (solid),
$c=0.1$ (dashed), $c=0.2$ (dotted), $c=0.3$ (dash-dotted),
and $c=0.4$ (dash-double-dotted). Here $\Gamma^+ = 0.15\,$meV
and $\Delta_0 = 24\sqrt{2}\,$meV.}
\label{f3}
\end{figure}
\begin{figure}
\caption{The real part of the optical conductivity $\sigma_1(\omega)$
in BCS theory with $\Gamma^+ = 0.15\,$meV at temperature
$T = 10\,$K for various values of the impurity parameter $c$,
namely Born limit ($c\to\infty$, solid), unitary limit ($c=0$,
dotted), $c=0.4$ (dashed), and $c=1.0$ (dash-dotted). The solid
gray curve is for comparison and gives the normal state.
The inset shows the same results on a different vertical scale.}
\label{f4}
\end{figure}
\begin{figure}
\caption{The microwave conductivity $\sigma_1(\omega,T)$ as a
function of $\omega$ for three different temperatures. The data
is the same as shown in Fig.~\ref{f1}. The open symbols are theory
for the pure limit, the solid gray symbols theory with some
impurity scattering additionally included, and the solid black
symbols are experiments. The squares are for $T=10\,$K, the
upward triangles for $T=15\,$K, and the down triangles for
$T=20\,$K.}
\label{f5}
\end{figure}

\begin{thebibliography}{99}
\bibitem{carb}J.P.~Carbotte, E.~Schachinger, and D.N.~Basov,
Nature (London) {\bf 401}, 354 (1999).
\bibitem{Bourges}Ph.\ Bourges, Y.\ Sidis, H.F.\ Fong, B.\ Keimer,
L.P.\ Regnault, J.\ Bossy, A.S.\ Ivanov, D.L.\ Milius, and
I.A.\ Aksay, in {\it High Temperature Superconductivity} ed.:
S.E.\ Barnes, {\it et al.}, CP483 American Institute of Physics,
Amsterdam (1999), p.\ 207-212.
\bibitem{schach1}E.~Schachinger and J.P.~Carbotte, \prb
{\bf 62}, 9054 (2000).
\bibitem{schach2}E.~Schachinger and J.P.~Carbotte, Physica C
{\bf 341-348} 79-82 (2000).
\bibitem{schach3}E.~Schachinger, J.P.~Carbotte, and D.N.~Basov,
Europhys. Lett. {\bf 54}, 380 (2001).
\bibitem{schach4}E.~Schachinger, J.P.~Carbotte, and F.~Marsiglio,
\prb {\bf 56}, 2738 (1997); {\it ibid.} {\bf 57} 7970 (1998).
\bibitem{millis}A.J.~Millis, H.~Monien, and D.~Pines,
\prb {\bf 42}, 167 (1990).
\bibitem{mont}P.~Monthoux and D.~Pines, \prb {\bf 47}, 6069
(1993).
\bibitem{dai}P.\ Dai, H.A.\ Mook, S.M.\ Hayden, G.\ Aeppli,
T.G.\ Perring, R.D.\ Hunt, and F.\ Do\u{g}an, Science {\bf 284},
1344 (1999).
\bibitem{nuss}M.C.~Nuss, P.M.~Mankiewich, N.L.~O'Malley,
and E.H.~Westwick, \prl {\bf 66}, 3305 (1991).
\bibitem{bonna}D.A.~Bonn, R.~Liang, T.M.~Risemann, D.J.~Baar,
D.C.~Morgan, K.~Zhang, P.~Dosanjh, T.L.~Duty, A.~McFarlane,
G.D.~Morris, J.H.~Brewer, W.N.~Hardy, C.~Kallen, and
A.J.~Berlinsky, \prb {\bf 47}, 11\ 314 (1993).
\bibitem{matsuk}M.~Matsukawa, T.~Mizukoshi, K.~Noto, and
Y.~Shiohara, \prb {\bf 53}, R6034 (1996).
\bibitem{Kee}H-Y.\ Kee, S.A.\ Kivelson, and G.\ Aeppli,
cond-mat/0110478 (unpublished).
\bibitem{hoss}A.~Hosseini, R.~Harris, S.~Kamal, P.~Dosanjh,
J.~Preston, R.~Liang, W.N.~Hardy, and D.A.~Bonn, \prb
{\bf 60}, 1349 (1999).
\bibitem{mars1}F.~Marsiglio, T.~Startseva, and J.P.~Carbotte,
Physics Lett. {\bf A245}, 172 (1998).
\bibitem{proh}M.~Prohammer and J.P.~Carbotte, \prb {\bf 43},
5370 (1991).
\bibitem{pan}S.H.~Pan, E.W.~Hudson, K.M.~Lang, H.~Eisaki,
S.~Uchida, and J.C.~Davis, Nature (London) {\bf 403},
746 (2000).
\bibitem{huds}E.W.~Hudson, S.H.~Pan, A.K.~Gupta, K.-W.~Ng, and
J.C.~Davis, Science {\bf 285}, 88 (1999).
\bibitem{frank}M.~Franz, C.~Kallin, A.J.~Berlinsky, and
M.I.~Salkola, \prb {\bf 56}, 7882 (1997).
\bibitem{ghosal}A.~Ghosal, M.~Randeria, and N.~Trivedi,
cond-mat/0004481 (unpublished).
\bibitem{franz} M.~Franz, C.~Kallin, and
A.J.~Berlinsky, \prb {\bf 54}, R6897 (1996).
\bibitem{zhu1}J.-X.~Zhu, C.S.~Ting, and Chia-Ren Hu, \prb {\bf 62},
6027 (2000).
\bibitem{atkinson1}W.A.~Atkinson, P.J.~Hirschfeld, and A.H.~MacDonald,
\prl {\bf 85}, 3922 (2000).
\bibitem{atkinson2}W.A.~Atkinson, P.J.~Hirschfeld, A.H.~MacDonald, and
K.~Ziegler, \prl {\bf 85}, 3926 (2000).
\bibitem{balatsky}A.V.~Balatsky, M.I.~Salkola, and A.~Rosengren,
\prb {\bf 51}, 15 547 (1995).
\bibitem{salkola}M.I.~Salkola, A.V.~Balatsky, and D.J.~Scalapino,
\prl {\bf 77}, 1841 (1996).
\bibitem{zhu2}J.-X.~Zhu and C.S.~Teng, \prb {\bf 63}, 020506 (2001).
\bibitem{hirsch1}P.J.~Hirschfeld and N.~Goldenfeld, \prb {\bf 48},
4219 (1993).
\bibitem{hirsch2}P.J.~Hirschfeld, W.D.~Putika, and D.~Scalapino,
\prb {\bf 50}, 10 250 (1994).
\bibitem{wu}Wen Chin Wu helped with the derivation of this result.
\bibitem{berlin}A.J.~Berlinsky, D.A.~Bonn, R.~Harris, and
C.~Kallin, \prb {\bf 61}, 9088 (2000).
\bibitem{schach5}E.\ Schachinger and J.P.\ Carbotte, \prb
{\bf 64}, 094501 (2001).
\end{thebibliography}
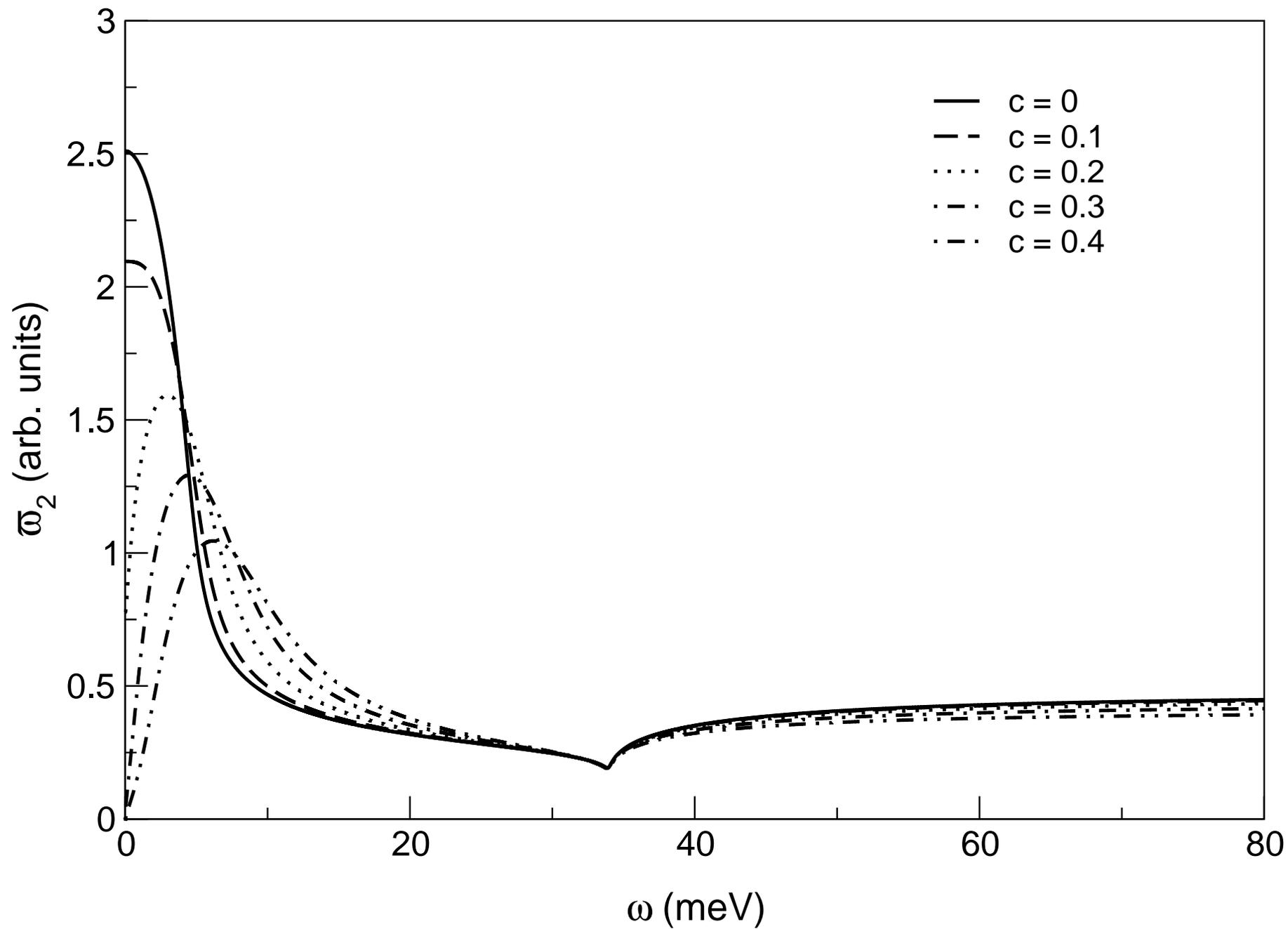
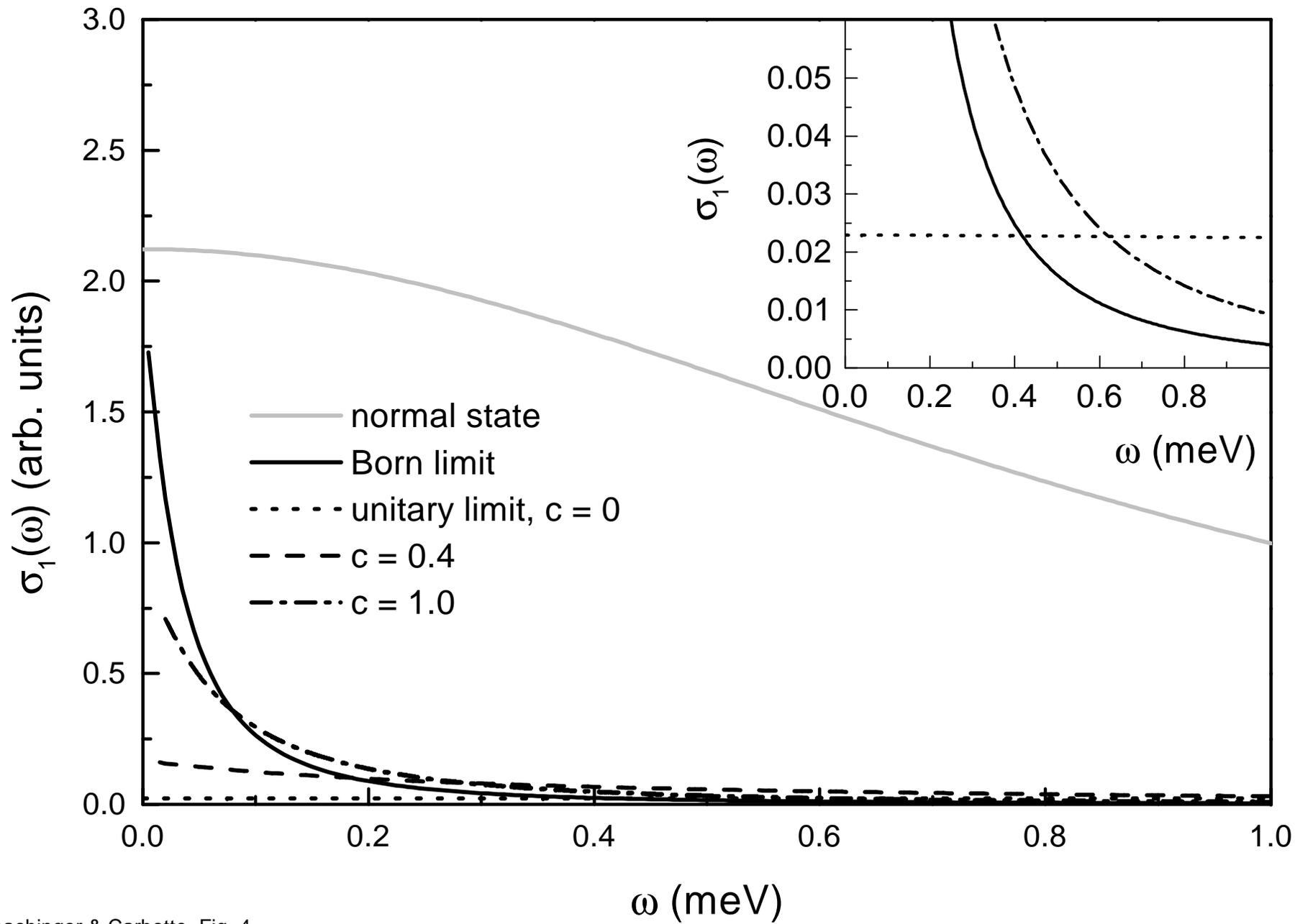
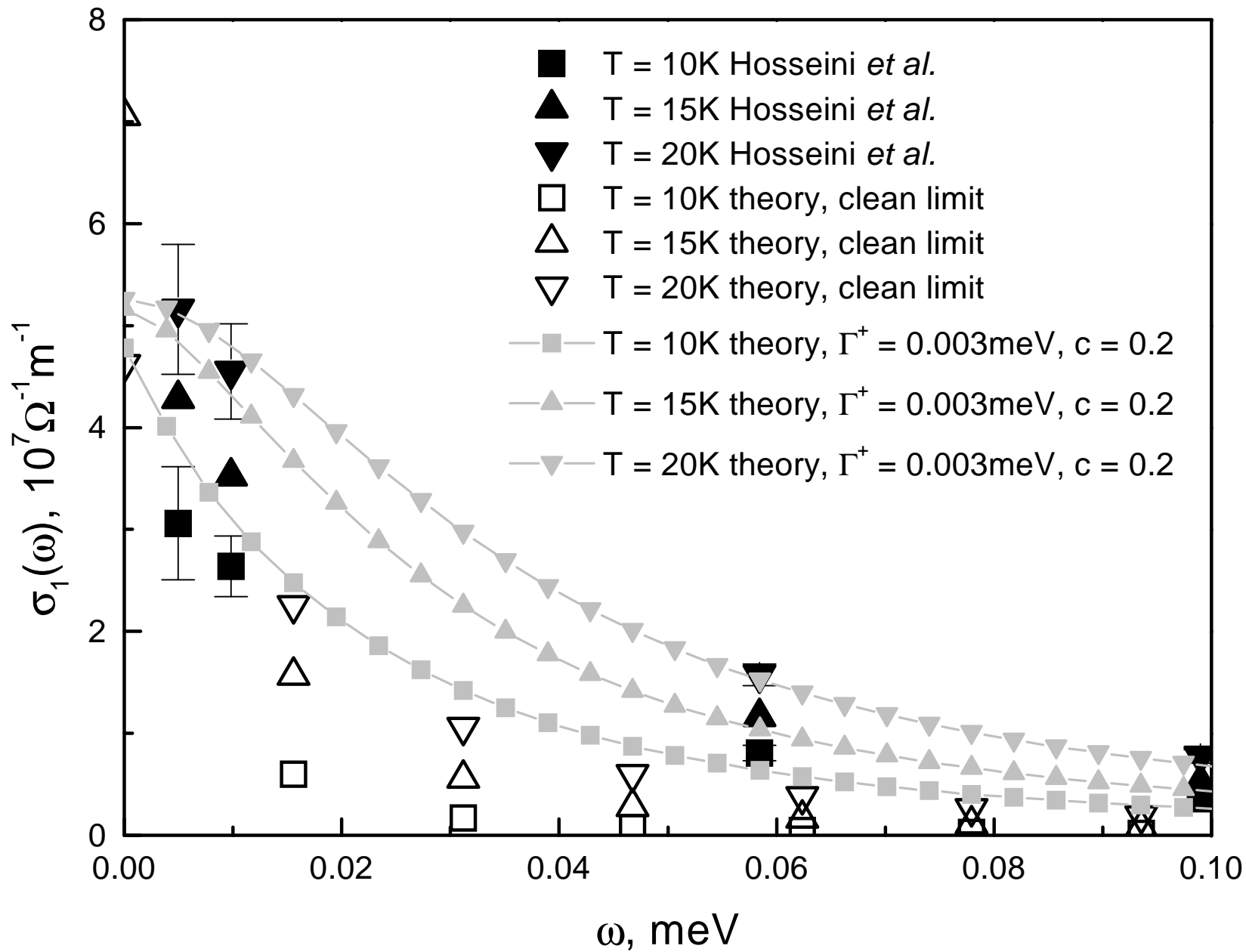
\end{document}